\documentclass[twocolumn]{aa}
\usepackage{csvsimple}

\DeclareUnicodeCharacter{2212}{-}

\newcommand{\cms}{\mbox{cm s$^{-1}~$}}
\newcommand{\kms}{\mbox{km s$^{-1}~$}}

\begin{document}
\titlerunning{EXPRES Transmission Spectroscopy of MASCARA-2 b}
\title{High-resolution Transmission Spectroscopy of MASCARA-2 b with EXPRES}

   \author{
    H. Jens Hoeijmakers\inst{1,2}
    \and
    Samuel H. C. Cabot\inst{3}
    \and
    Lily Zhao\inst{3}
    \and
    Lars A. Buchhave\inst{4}
    \and
    Ren\'e Tronsgaard\inst{4}
    \and
    Daniel Kitzmann\inst{2}
    \and
    Simon L. Grimm\inst{2}
    \and
    Heather M. Cegla\inst{1}
    \and
    Vincent Bourrier\inst{1}
    \and
    David Ehrenreich\inst{1}
    \and
    Kevin Heng\inst{2,5}
    \and
    Christophe Lovis\inst{1}
    \and
    Debra A. Fischer\inst{3}
    }

   \institute{
    Observatoire de Gen\`eve, Chemin des Maillettes 51, 1290, Versoix, Switzerland
        \and
    Center for Space and Habitability, Universit\"at Bern, Gesellschaftsstrasse 6, 3012 Bern, Switzerland
        \and
    Yale University, 52 Hillhouse, New Haven, CT 06511, USA
        \and
    DTU Space, National Space Institute, Technical University of Denmark, Elektrovej 328, DK-2800 Kgs. Lyngby, Denmark
        \and
    University of Warwick, Department of Physics, Astronomy \& Astrophysics Group, Coventry CV4 7AL, United Kingdom
             }
\date{Received January 2, 2020; accepted April 17, 2020}

\abstract{
We report detections of atomic species in the atmosphere of MASCARA-2 b, using the first transit observations obtained with the newly commissioned EXPRES spectrograph.
EXPRES is a highly stabilised optical echelle spectrograph, designed to detect stellar reflex motions with amplitudes down to 30 cm/s, and was recently deployed at the Lowell Discovery Telescope. By analysing the transmission spectrum of the ultra-hot Jupiter MASCARA-2 b using the cross-correlation method, we confirm previous detections of \ion{Fe}{I}, \ion{Fe}{II} and \ion{Na}{I}, which likely originate in the upper regions of the inflated atmosphere. In addition, we  report significant detections of \ion{Mg}{I} and \ion{Cr}{II}. The absorption strengths change slightly with time, possibly indicating different temperatures and chemistry in the day-side and night-side terminators. Using the effective stellar line-shape variation induced by the transiting planet, we constrain the projected spin-orbit misalignment of the system to $1.6\pm3.1$ degrees, consistent with an aligned orbit.
We demonstrate that EXPRES joins a suite of instruments capable of phase-resolved spectroscopy of exoplanet atmospheres.
}

\keywords{}

\maketitle

\section{Introduction} \label{sec:intro}

The emergence of a new generation of spectrographs opens avenues for detecting and characterizing exoplanets and their atmospheres with unprecedented fidelity. Exoplanets can be detected with high-resolution spectrographs due to their gravitational interaction with the host-star, causing a periodic Doppler shift in the observed stellar spectrum as the planet orbits. These measurements yield a lower limit to the mass of the planet or the mass itself if the planet orbits in an aligned plane that carries it through transit as seen from Earth. One may constrain the composition of a transiting planet by also measuring its radius, which then yields the mean density. Recent years have seen extensive efforts to develop a new generation of environmentally-controlled high-resolution spectrographs that provide ever increasing radial velocity (RV) precision and stability, to allow for more precise mass measurements, and for the detection of less massive planets further away from their host star. Indeed, several purpose-built instruments have come online in recent years. ESPRESSO \citep{PePe2013} was commissioned at the Very Large Telescope (VLT) in December 2017. HARPS-N/GIANO \citep{Cosentino2012,Claudi2017, Oliva2018} at Telescopio Nazionale Galileo, CARMENES \citep{Quirrenbach2010} at Calar Alto Observatory and SPIRou \citep{Thibault2012} at the Canada France Hawaii Telescope provide high-resolution optical and NIR coverage. NIRPS \citep{Wildi2017} at the ESO's 3.6-m telescope at La Silla and HARPS3 \citep{Thompson2016} at the Isaac Newton Telescope are currently under development.

The EXtreme PREcision Spectrograph (EXPRES) was commissioned at the Lowell Observatory 4.3-m Lowell Discovery Telescope \citep[LDT,][]{Levine2012} in 2018. EXPRES is a vacuum-stabilized, fiber-fed, $R \sim 140,000$, optical spectrograph with wavelength calibration from a Menlo Systems laser frequency comb (LFC). It is optimized for wavelengths between 380 and 680 nm.
The design goal of EXPRES was a RV precision of $\sim 30$ \cms on bright (V $<$ 8) main sequence stars. Observations with the LFC demonstrate an instrumental stability of 7 \cms and formal errors of about 25 \cms for an SNR of 250 when observing bright stars \citep{Blackman2019,Petersburg2020}.

High resolution spectrographs are able to directly probe spectral features in the atmospheres of exoplanets \citep{Snellen2008}. The high spectral resolution allows for individual spectral lines in the planet's spectrum to be resolved and provides measurements of the line-shape and depth \citep[e.g.][]{Redfield2008, Wyttenbach2015, Jensen2012, Khalafinejad2017, Wyttenbach2017, Allart2019, Seidel2019}. The orbital velocity of hot-Jupiters is often in excess of 100 \kms. At $R\sim10^5$, this motion is resolved in time-series exposures of a several-hour transit. Separating the spectrum of the planet from the host star using this Doppler shift was first applied to detect absorption by atmospheric CO in the infra-red transmission spectrum of HD 209458 b \citep{Snellen2010}. Later efforts employed this technique to detect the thermal emission of both transiting and non-transiting hot-Jupiters, resulting in detections of CO, H$_2$O, CH$_4$ and HCN \citep[e.g.][]{Brogi2012, Birkby2013,Lockwood2014,Brogi2016,Piskorz2016,Birkby2017,Hawker2018, Cabot2019, Guilluy2019, Flagg2019}. The technique has also been applied at optical wavelengths to detect TiO in the day-side spectrum of WASP-33 b using the HDS/Subaru instrument \citep{Nugroho2017}, and atomic metal lines in the transmission spectrum of KELT-9 b \citep{Hoeijmakers2018, Hoeijmakers2019}  using HARPS-N/$TNG$ and PEPSI/$LBT$ \citep{Cauley2019}. Ongoing improvements in resolution, stability, and wavelength calibration will enable more detailed detections of chemistry and atmospheric dynamics for a multitude of exoplanet systems \citep{Heng2015,Crossfield2015, Madhusudhan2016, Birkby2018, Triaud2018, Wright2018}. As such, stabilized, fiber-fed high-resolution spectrographs have a complementary dual-use in the discovery of exoplanets as well as the characterization of their atmospheres.

MASCARA-2 b/Kelt-20 b (hereafter M2) was independently discovered by the Multi-site All-Sky CAmeRA (MASCARA) \citep{Talens2018} and the Kilo-degree Extremely Little Telescope (KELT) \citep{Lund2017}, which both survey bright stars for signs of periodically transiting exoplanets. The planet is a hot-Jupiter, transiting the bright ($m_V = 7.6$) A2 main sequence star HD 185603 in a 3.47 day orbit. The orbit of the planet is aligned, which is not common for this type of star \citep{Winn2010,Schlaufman2010,Albrecht2012}. Strong irradiation from its host star gives M2 b a high equilibrium temperature of $T_{eq} \sim 2260$ K, placing it in the extreme class of ultra-hot Jupiters. Several such planets have been studied to date, most notably KELT-9 b, which, with $T_{eq} = 4050$ K is the hottest known planet around a main-sequence star. In light of previous detections of atomic metals in KELT-9 b \citep{Hoeijmakers2018, Hoeijmakers2019}, atmospheric studies of other ultra-hot Juptiers warrant searches for vaporized metals and ions.

The transmission spectrum of M2 has been observed extensively with the HARPS-N and CARMENES spectrographs, leading to detections of two Balmer lines of hydrogen, the \ion{Na}{I} D-lines, the \ion{Ca}{II} infra-red triplet, \ion{Fe}{I}, \ion{Fe}{II} and tentative evidence for an \ion{Mg}{I} line at 517.268 \citep{Casasayas2018,Casasayas2019,Nugroho2020,Stangret2020}. These detections are indicative of the high temperatures prevalent at the day-side and terminator regions. The detection of strong hydrogen absorption may further be indicative of an extended or evaporating atmosphere \citep{Yan2018, Turner2019}.

This paper presents the results of one night of transit observations of M2 with the EXPRES instrument, which constitute the first application of EXPRES for the purpose of exoplanet transit transmission spectroscopy. With these observations, we aim to demonstrate that EXPRES offers significant potential for atmospheric characterization in addition to its main purpose of RV monitoring for exoplanet discovery. The paper is organized as follows: in section \ref{sec:obs} we describe the observations and present our analysis of the transmission spectra. In section \ref{sec:res}, we discuss our results, including analysis of the Doppler shadow induced by the  Rossiter-McLaughlin effect. We detect spectroscopic signatures of the atmosphere of M2 by confirming the presence of \ion{Fe}{I} and \ion{Fe}{II} \citep{Casasayas2019,Nugroho2020,Stangret2020}, and presenting strong evidence for  \ion{Mg}{I}, \ion{Na}{I} and \ion{Cr}{II}. We summarize our results in Section \ref{sec:con}.

\section{Transit observations of MASCARA-2 b}
\label{sec:obs}
A single transit of M2 was observed with the LDT during the night of June 1, 2018.  The night was clear with an average seeing of approximately 1."2.  The observations lasted from 09:03 to 14:23 UTC, with a total of 68 exposures of the system, of which 51 were obtained in-transit (we include ingress/egress in our in-transit sample). The signal to noise ratio for these exposures reached approximately 40-50 near 5000 \AA throughout the night. Most exposures were approximately 200 seconds, while 15-minute exposures at the start of the night provided high S/N out-of-transit spectra.

\begin{figure*}
   \centering
   \includegraphics[width=1.0\linewidth,angle=0]{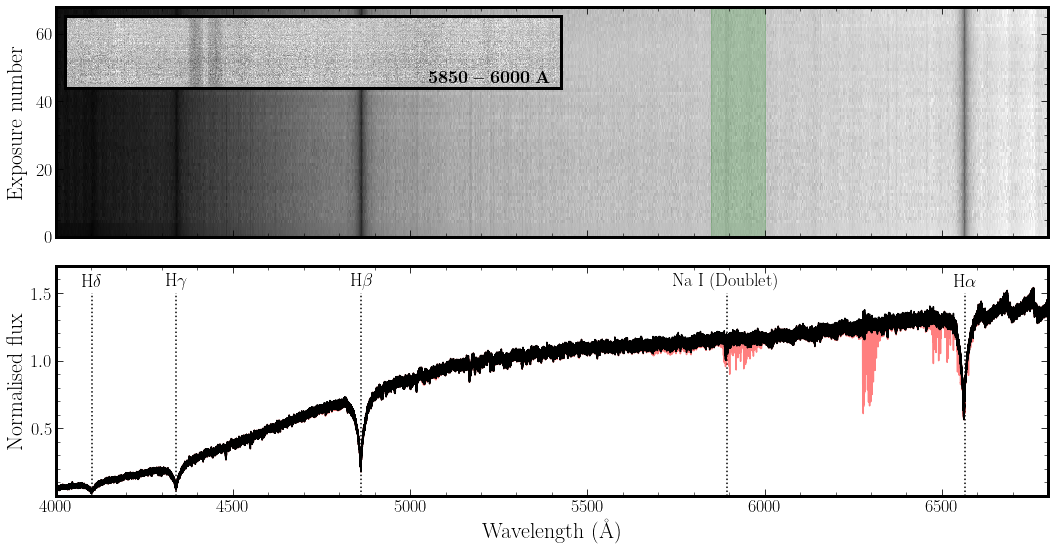}
   \includegraphics[width=1.0\linewidth,angle=0]{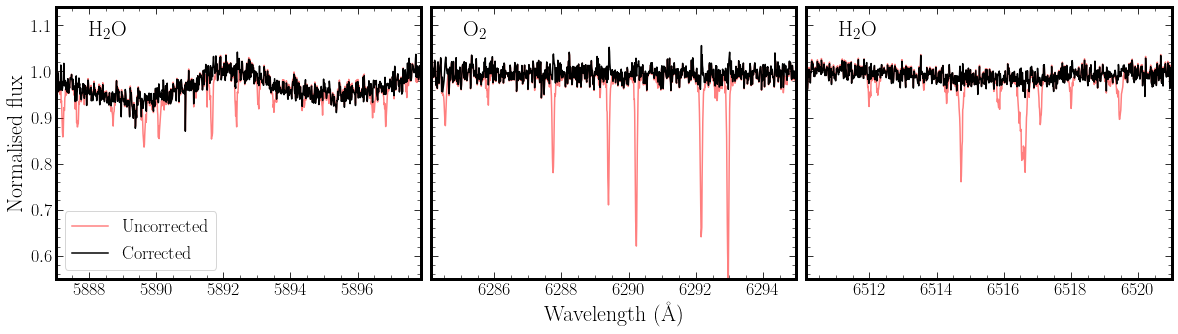}
      \caption{
      Spectra obtained on June 1, 2018 with EXPRES. {\it Top Panel}: Time-series, high-resolution spectra cleaned and corrected for tellurics. The orders are stitched together for visualization, but are analyzed independently during cross-correlation. Common features such as stellar lines appear as dark columns. {\it Inlet}: Zoom-in of the time-series spectra between 5850-6000 \AA\ (marked by the green band), revealing the broadened Na doublet. {\it Middle Panel}: Example 1-dimensional spectrum from the time-series, before (red) and after (black) telluric correction. The features around 6700 \AA\ are an artifact of order stitching. Some strong stellar lines, including the Balmer series and the Sodium D lines are marked. {\it Bottom Panels}: Zoom-in of the example 1-dimensional spectrum, demonstrating the effect telluric correction in strongly contaminated regions (due to H$_2$O, O$_2$ and H$_2$O, from left to right).}
         \label{fig:vis}
   \end{figure*}

Wavelength calibration is carried out with a Menlo Systems Laser Frequency Comb as well as a ThAr line-lamp.  Exposures are taken through the science fiber and are interspersed with science observations approximately every 30 minutes. A tune-able LED is used to take flat-fields. Sets of zero-second bias frames are taken before and after the observing night, and dark current is corrected using the CCD overscan regions. We perform extractions with the \texttt{RePack}\footnote{Written by Lars Buchhave} code, adapted from a version purposed for use with HARPS-N data products \citep{Fischer2016}. We also divide spectra by a blaze-function derived from flat-field calibrations. The result is a pipeline-reduced spectral time series in Earth's rest-frame, shown in the top panel of Figure \ref{fig:vis}.

Our subsequent analysis follows the approach of \citet{Hoeijmakers2019} that resulted in the detections of iron and titanium in the atmosphere of KELT-9 b, using a similar sequence of observations by the HARPS-N spectrograph. The cross-correlation procedure discussed below differs from the approach used by \citet{Casasayas2018, Casasayas2019}, which was based on the direct analysis of individual absorption lines in a single, co-added transmission spectrum.

\subsection{Telluric Correction and Detrending}
Absorption lines from the stellar photosphere as well as the Earth's atmosphere (telluric lines) are the most dominant spectral features in the observed spectra. We model the telluric contamination in each spectrum with the \texttt{Molecfit} package \citep{Smette2015}. \texttt{Molecfit} processes an observed spectrum along with ambient weather, time, and location information to model the structure of the Earth's atmosphere during an observation and to fit a line-by-line radiative transfer model to the spectrum. We use similar initial parameters to those used by \citet{Allart2017}, such as the atmospheric profile, degree of the continuum and wavelength polynomial fits, and resolution fit kernel. We chose more relaxed convergence tolerances of 10$^{-5}$, which improved runtime and the quality of fit. Because M2 is a fast rotator, most stellar lines are significantly broadened and blended to form a quasi-continuum, minimizing their effect on the fit of individual telluric lines. We fit each spectrum in the time-series individually and divide the spectrum by the fit to remove tellurics (Figure \ref{fig:vis}, middle and bottom panels).

Our analysis uses spectra in the wavelength range 4000-6800 \AA. Bluer wavelengths are subject to very low instrumental throughput, and redder wavelengths suffer severe telluric contamination that is more difficult to model while covering comparatively few atomic metal lines. This leaves 64 out of 88 orders, which are treated separately until combining their individual cross-correlation functions at the end of the analysis. For computational reasons, we re-sample all of the spectra onto a common wavelength grid of 0.01 \AA\ spacing. We identify bad pixels (e.g. due to cosmic ray hits) by selecting 5$\sigma$ outliers in a sliding 500 pixel window, and setting them to the mean of the values in the window. We shift each spectrum into the rest-frame of the star, taking into account the -21.07 \kms systemic velocity (denoted $V_{\rm sys}$) \citep{Talens2018} and the Barycentric Earth Radial Velocity (BERV) correction \citep[calculated using the \texttt{Astropy} package, see][]{astropy:2013, astropy:2018}. The BERV changes by $\lesssim 1$ \kms over the course of the transit. We do not correct for stellar reflex motion. This effect is negligible because M2 has broad absorption lines due to its fast rotation. We note the presence of excess absorption in the cores of the Na doublet which we attribute to stationary ISM absorption, as seen in previous analyses of M2 \citep{Casasayas2018, Casasayas2019}. As in these studies, neglecting the stellar reflex correction helps ensure the excess absorption cancels during division by the master out-of-transit spectrum (see below). The 2.5$\%$ of pixels on either end of each order are masked from cross-correlation. This step removes edge artifacts from shifting to the stellar rest-frame, as well as pixels with low flux located at the edges of the blaze function.

We co-add all out-of-transit spectra $\{f_t\}_{t\in t_{\rm out}}$ to obtain a master spectrum of the star $F_{\rm out}$.  We subsequently compute individual transmission spectra $\{\widetilde r_t\}_{t\in t_{\rm in}}$ by dividing each in-transit spectrum $\{f_t\}_{t\in t_{\rm in}}$ by $F_{\rm out}$ \citep{Wyttenbach2015, Allart2017, Hoeijmakers2018, Casasayas2018, Casasayas2019}. Finally, we apply a high-pass Gaussian filter with 75 pixel standard-deviation and subtract from the pixels in each wavelength bin their time-average, removing broad-band variations \citep{Hoeijmakers2018b}.

\subsection{Cross-correlation}
Atomic or molecular species present in the exoplanet atmosphere may cause thousands of individual absorption lines in the atmospheric transmission spectrum. Cross-correlating the spectra with a model template spectrum combines the contributions of all these lines to reduce the photon-noise and yield significant detections of these elements \citep{Snellen2010}. We use high-resolution model spectra for \ion{Na}{I}, \ion{Mg}{I}, \ion{Sc}{I}, \ion{Sc}{II}, \ion{Ti}{I}, \ion{Ti}{II}, \ion{Cr}{I}, \ion{Cr}{II}, \ion{Fe}{I}, \ion{Fe}{II} and \ion{Y}{II} using opacities computed with \texttt{HELIOS-K} \citep{Grimm2015} and equilibrium chemistry at $T = 4,000$ K with \texttt{FastChem} \citep{Stock2018}. These templates resulted in detections in the analysis of the transmission spectrum of KELT-9 b by \citet{Hoeijmakers2019}, who provide a detailed description of their construction. These templates are publicly accessible via the CDS \footnote{\url{https://cdsarc.unistra.fr/viz-bin/cat?J/A+A/627/A165}}.

To compute the cross-correlation, we apply Doppler shifts of -500 to 500 \kms to the template in increments of 1.0 \kms. At each velocity $v$, and for a given transmission spectrum $\widetilde r_t$, we take the weighted average of values in the spectrum. The weights are a product of the Doppler-shifted template, which is zero everywhere except at locations of line transitions, and the inverse of each wavelength bin's time-variance, which reduces the contribution of noisy pixels \citep{Brogi2016}. This procedure yields a cross-correlation function (CCF) as a function of velocity and time, which contains the average line-strength of features in the transmission spectrum \citep{Hoeijmakers2019}.

\subsection{Rossiter-McLaughlin effect}
\label{sec:rm}
The orbit of M2 b is aligned \citep{Lund2017,Talens2018}, meaning that the planet obscures the blue-shifted part of the star at the start of the transit, and moves to the red-shifted part towards egress. This passage across the stellar disk changes the shapes of the disk-integrated, rotation-broadened stellar absorption lines, which gives rise to the Rossiter-McLaughlin (RM) effect seen in the RV measurements of the star during transit \citep{Ohta2005}. For a fast rotator like M2, broadened spectral features decrease the precision of RV measurements, making the RM effect difficult to resolve. However, when dividing the mean out-of-transit spectrum from each of the exposures, the line-deformation causes the stellar lines to be over-corrected at the instantaneous velocity of the obscured part of the stellar disk. This effect is known as the "Doppler shadow" that is cast by the planet as it progresses through transit \citep{CC2010}. The centroid velocity $v_*(t)$ of the obscured stellar surface depends on the projected location of the planet with respect to the stellar spin axis, which changes over time as the planet progresses through transit, and is described by the following analytical expressions \citep{CC2010,Bourrier2015}, ignoring differential rotation of the stellar surface and convective blueshift \citep{Cegla2016}:

\begin{equation}
    v_*(t) = x_{\perp}(t) v_{\textrm{eq}} \sin i_* \,
\label{eqn:rm}
\end{equation}

where $v_{\textrm{eq}} \sin i_*$ is the projected equatorial rotation velocity of the star, and $x_{\perp}$ is the orthogonal distance of the obscured region to the projected spin axis:

\begin{equation}
    x_{\perp}(t) = x_p(t) \cos (\lambda) - y_p(t) \sin (\lambda) \,
\end{equation}

where $\lambda$ is the spin-orbit misalignment and $x_p(t)$ and $y_p(t)$ are the coordinates of the planet:

\begin{equation}
    x_p(t) = \frac{a}{R_*} \sin (2 \pi \phi) \,
\end{equation}

\begin{equation}
   y_p(t) = −\frac{a}{R_*} \cos (2 \pi \phi) \cos (i_p)   \,
\label{eqn:yp}
\end{equation}

with $a$ the semi-major axis, $R_*$ the radius of the star, $i_p$ the orbital inclination, $\phi$ the orbital phase at time $t$ and the orbital inclination $i_p$.

The Doppler shadow is retrieved by cross-correlating the transmission spectra $\{\widetilde r_t\}_{t\in t_{\rm in}}$ with a model template of the stellar spectrum. This template consists of a PHOENIX photosphere model spectrum at a temperature of 9000 K and solar metallicity, obtained from the online PHOENIX library \citep{Husser2013}. The baseline of the template is subtracted via a high-pass filter so that it contains only the absorption lines with relative depths corresponding to the relative absorption of continuum radiation at each wavelength (i.e. continuum normalization).

\begin{figure*}
   \centering
   \includegraphics[width=1.0\linewidth]{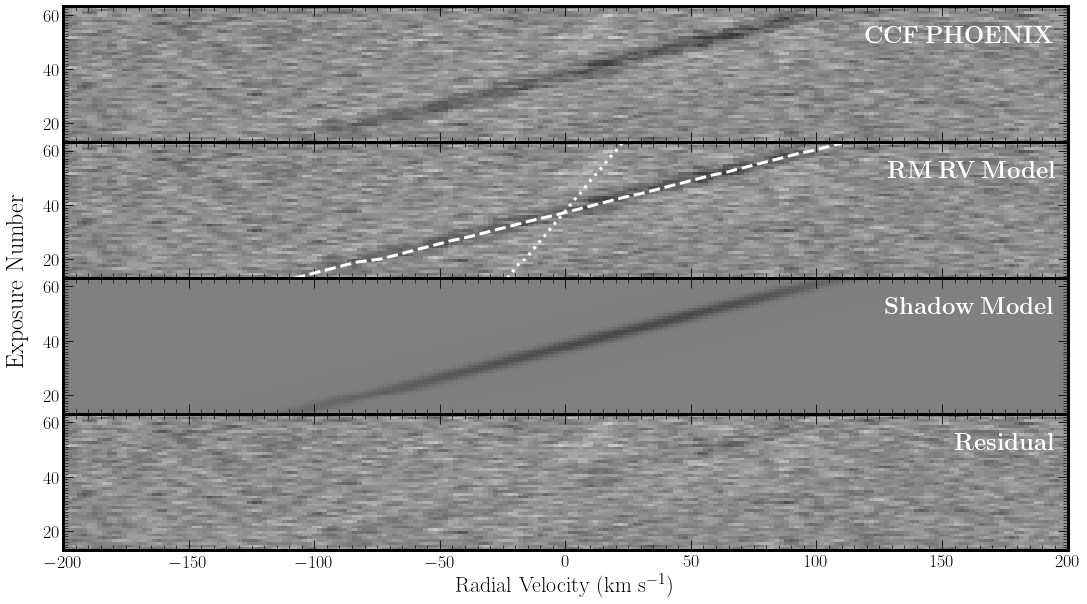}
      \caption{Removal of the Doppler shadow caused by the transiting planet, from start of ingress to end of egress. {\it Top panel}: The cross-correlation of the exposures in-transit with a template constructed from a PHOENIX stellar photosphere model. {\it Second panel}: Same as top panel, annotated with the predicted RV of the occulted stellar region responsible for the Doppler shadow (dashed line). Also shown is the expected RV of the planet (dotted line) expected from the system parameters by \citep{Lund2017,Talens2018}. {\it Third Panel}: The model of the Doppler shadow, obtained by fitting a Gaussian profile to the shadow feature in each of the in-transit cross-correlation functions. {\it Bottom panel}: Residuals after subtraction.}
     \label{fig:dopplershadow}
   \end{figure*}

The resulting two-dimensional CCF contains the signature of the Doppler shadow (see the top panel in Figure \ref{fig:dopplershadow}), from which the misalignment $\lambda$ between the orbital plane of the planet and the projected stellar spin axis can be derived as is done when applying Doppler tomography \citep{CC2010} or the ``Reloaded RM-effect`` \citep{Cegla2016}. The shadow also overlaps with the expected RV of the planet, so it needs to be corrected before the signature of the planet atmosphere may be isolated (see the second panel, "RM RV Model", in Figure \ref{fig:dopplershadow}; the local obstructed RV overlaps the planetary RV, given by dashed and dotted lines respectively). We construct a model of the two-dimensional cross-correlation residual from the parameters of the Gaussian fits used to measure $\lambda$, following the approach by \citet{Hoeijmakers2018}. This model is scaled to minimize the sum of the squared residual when subtracting it from cross-correlation functions with other the templates used in this analysis (see third and fourth panels, "Shadow Model" and "Residual", in Figure \ref{fig:dopplershadow}). The Doppler shadow shows small variations in strength throughout the transit; darker regions are roughly correlated with higher exposure S/N. However, our empirical model is agnostic to the origin of these variations, and successfully removes the shadow to isolate the atmospheric absorption. The overlap region between the atmospheric absorption trail and the Doppler shadow may affect the co-added absorption signal; however, as described by \citet{Hoeijmakers2018}, the fit of Doppler shadow model ignores the overlap region and enforces smoothly varying model parameters, which helps to preserve information at velocities for which the Doppler shadow and the planet absorption features overlap. This is possible because range of radial velocities spanned by the Doppler shadow differs significantly different from those of the planet atmosphere, due to the fast rotation of the host star.

\subsection{The time-averaged CCF}

The planet's apparent RV is described by:
\begin{equation}
    v(t) = K_{\rm p} \sin (2\pi\phi(t)) + V_{\rm sys}
\end{equation}
where $K_{\rm p}$ is the planetary semi-amplitude. Note that since we shifted all spectra to the stellar rest-frame, the $V_{\rm sys}$ term above is set to 0.0 \kms. The orbital period $P=3.474119$ days and reference mid-transit time $T_0=57909.0875$ MJD \citep{Talens2018} determine the orbital phase, $\phi(t)$ at the time of each exposure. We sample potential values for $K_{\rm p}$ from a grid ranging from 0-300 \kms in steps of 1.0 \kms and shift each CCF by $-K_{\rm p} \sin (2\pi\phi(t))$. At the true value of $K_{\rm p}$, the CCFs are shifted into the rest from of the planet, and may be optimally co-added to yield the time-average of the CCFs as a function of the systemic velocity $V_{\rm sys}$. The planet signal is expected to occur at 0.0 \kms because the spectra have been shifted to the stellar rest-frame. In this way, we construct a two-dimensional map of the co-added cross-correlation signal at different combinations of $K_{\rm p}$ and $V_{\rm sys}$, which has been termed the $K_pV_{\rm sys}$ diagram \citep{Brogi2012}. A statistically significant signal at the correct combination of $K_p$ and $V_{\rm sys}$ confirms the presence of the model species in the atmosphere of the planet \citep{Brogi2012}. Additionally, we can examine a single row in this map that corresponds to any particular choice of $K_{\rm p}$. The peak in this co-added CCF corresponds to the weighted mean of the depths of the absorption lines in the chosen rest-frame \citep{Pino2018, Hoeijmakers2019}. The orbital period and semi-major axis of M2 \citep{Talens2018} and previous atmospheric measurements \citep{Casasayas2019,Nugroho2020,Stangret2020} place $K_{\rm p}$ roughly between $160-190$ \kms.

\begin{figure}
   \centering
   \includegraphics[width=\linewidth,angle=0]{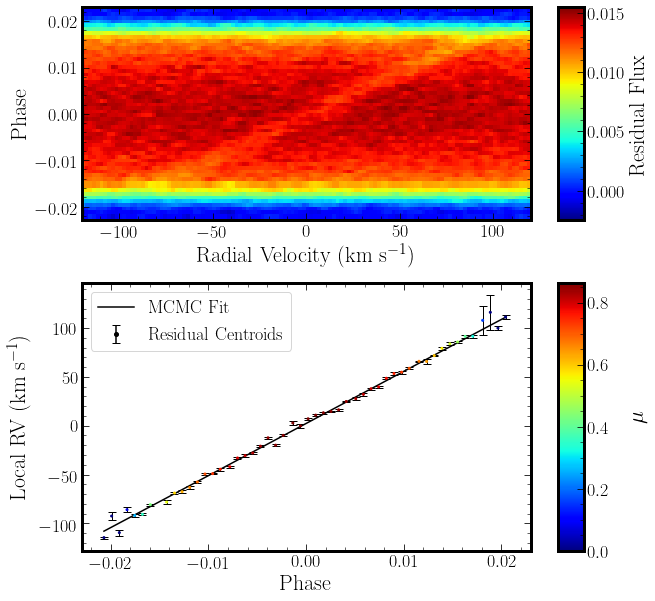}
      \caption{{\it Top Panel:} The two-dimensional CCF obtained using the PHOENIX stellar template, normalized by the transit light-curve. {\it Bottom Panel:} The functional form of the time-dependent centroid velocity of the Doppler shadow is fit using an MCMC optimizer, providing posterior distributions of the system parameters (see Fig. \ref{fig:rmmcmc}).}
\label{fig:RMreloaded}
\end{figure}

\begin{figure*}
   \centering
   \includegraphics[width=0.9\linewidth,angle=0]{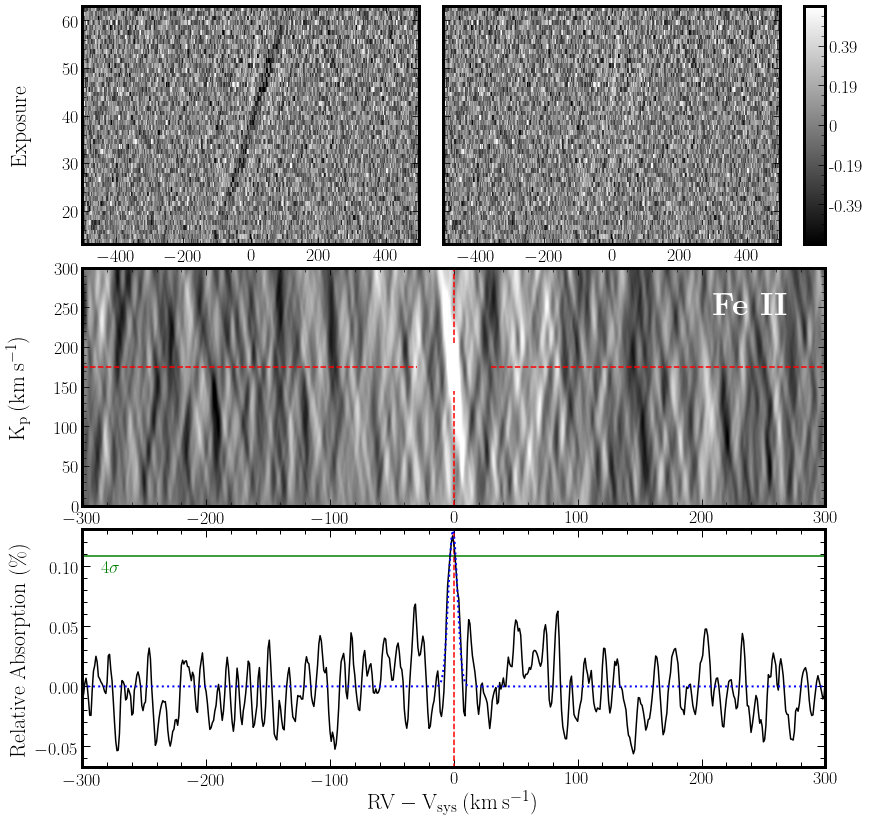}
      \caption{Cross-correlation procedure and detection of \ion{Fe}{II} in the transmission spectrum of MASCARA-2 b. {\it Upper panels}: two-dimensional cross-correlation with the  \ion{Fe}{II} template before (left) and after (right) correction of the Doppler shadow. {\it Middle panel}: $K_pV_{\textrm{sys}}$ diagram. The red dashed lines indicate the expected location of the atmospheric signal. Because the data are shifted by the systemic velocity and Barycentric Earth Radial Velocity, the signal lies at approximately 0 along the x-axis. Bottom panel: Cross-correlation function co-added in the rest-frame of the planet. The green line marks $4\times$ the standard deviation of the CCF. An enhancement is detected at the rest-frame velocity of the planet, which is modeled with a Gaussian profile (blue dotted line).}
\label{fig:result}
\end{figure*}

\section{Results}
\label{sec:res}

\subsection{Spin-orbit misalignment}

By making use of the fact that the RV variation of the Doppler shadow traces the obscured area of the stellar disk as the planet moves through transit, the system architecture can be derived \citep{CC2010}. To this end, we normalize the two-dimensional CCF by the expected flux decrease during the planet transit (i.e. the transit light-curve) and fit the centroid position of the correlation excess with a Gaussian profile in the cross-correlation function of each exposure, following \citet{Cegla2016} (see Fig. \ref{fig:RMreloaded}). We take the centroid $v^{obs}_*(t)$ and its uncertainty $\sigma^{obs}_{v}(t)$ as our measurement of the average RV of the occulted stellar surface. That is, $v^{obs}_*(t)$ represents the measurement of $v_*(t)$ which is defined in Equation \ref{eqn:rm}. We assume a standard Gaussian log-likelihood function,
\begin{equation}
    \mathcal{L} \propto -\frac{1}{2}\sum_t{\Big(\frac{v^{obs}_*(t)-\hat{v}_*(t)}{\sigma^{obs}_{v}(t)}\Big)^2}
\end{equation}
where $\hat{v}_*(t)$ is the model-predicted value of $v_*(t)$. We subsequently perform a Markov Chain Monte Carlo (MCMC) analysis of the Doppler shadow to constrain orbital parameters of the system.

We assume Gaussian priors on each of the orbital parameters, with mean and standard deviation as reported by \citet{Talens2018} (see Table \ref{tab:rmtab}). Our model contains 5 parameters: scaled semi-major axis $a/R_*$, projected obliquity $\lambda$, orbital inclination $i_p$, projected stellar rotation speed $v \sin i_*$, and a systematic RV offset to account for errors in the systemic velocity. We place a Gaussian prior on the offset with mean and standard deviation of 0.0 and 5.0 \kms respectively. The MCMC is performed with the \texttt{emcee} package \citep{Foreman-Mackey2013}, using 20 independent chains, each taking $1\times10^5$ steps. Visual inspection of the chains suggests a burn-in time of $\sim$ 200 steps, and an auto-correlation length of $\sim$ 3 steps. We exclude the first 5000 steps of each chain and nine out of every ten remaining steps, and subsequently merge the chains, leaving $>10^5$ independent samples of the posterior. Correlation diagrams and histograms of each parameter are shown in Figure \ref{fig:rmmcmc}, and the median values are quoted in Table \ref{tab:rmtab}. Our results are generally consistent with \citet{Talens2018} and \citet{Lund2017}. The marginalized distributions reduce the 1$\sigma$ uncertainty in $v\sin i_*$ to  $<1$ \kms, slightly improve constraints on $\lambda$, and constrain an absolute RV offset to within a few \kms. Distributions for $a/R_*$ and $i_p$ are dominated by the choice of priors, which are provided by the already tight constraints derived from transit photometry \citep{Talens2018}. Our best-fit parameters, in combination with Equations \ref{eqn:rm}-\ref{eqn:yp}, and the orbital phase of each exposure, trace the path of the Doppler shadow (second panel, "RM RV Model", Figure \ref{fig:dopplershadow}). Importantly, we note a degeneracy between RV offset and $\lambda$, which highlights the need for precise measurements of $V_{\rm sys}$ when fitting for the spin-orbit misalignment. However, it is difficult to accurately measure $V_{\rm sys}$ for fast-rotators like M2, and literature estimates vary by several \kms.

\subsection{Transmission spectrum and cross-correlation}
We present new detections of \ion{Cr}{II} (4.1$\sigma$) and \ion{Mg}{I} (4.0$\sigma$) in the atmosphere of M2 b, in addition to confirmations of \ion{Fe}{I} (4.7$\sigma$), \ion{Fe}{II} (4.8$\sigma$) and \ion{Na}{I} (4.4$\sigma$) \citep{Casasayas2019,Nugroho2020,Stangret2020}. The cross-correlation procedure and detection of \ion{Fe}{II} is shown in Figure \ref{fig:result}; the same plots for the remaining species are in the Appendix.

Several metrics have previously been used to determine detection significance in cross-correlation functions, including the amplitude of the CCF peak relative to the standard deviation of noise in the baseline of the CCF \citep{Brogi2012}, a Welch t-Test comparing distributions in the two-dimensional CCF \citep{Birkby2017}, and the uncertainty in a Gaussian fit to the CCF feature \citep{Hoeijmakers2019}. The significances determined in the present work represent the false-alarm probability (FAP) of detecting a comparable CCF enhancement by chance. We discuss FAP calculation in the following section. Summary statistics for each detected species, including the CCF peak, Gaussian fit, and FAP are listed in Table \ref{tab:amptab}.

\begin{table*}
  \centering
  \begin{tabular}{ccccccccc}
\hline
Species & CCF$_{\rm max}$ ($\%$) & S/N & $v_{\rm CCF}$ (km s$^{-1}$) & ${A}$ ($\%$) & ${\mu}$ (km s$^{-1}$) & ${\text{FWHM}}$ (km s$^{-1}$) & $p_{\rm FAP}$ & $\sigma_{\rm FAP}$ \\
\hline
\hline
\ion{Fe}{I} & 0.015 & 3.45 & -2 & 0.013$\pm$0.002 & -4.81$\pm$0.72 & 12.39$\pm$1.69 & 1.19e-06 & 4.72\\
\ion{Fe}{II} & 0.125 & 4.60 & -1 & 0.130$\pm$0.011 & -0.75$\pm$0.37 & 8.54$\pm$0.87 & 9.72e-07 & 4.76\\
\ion{Cr}{II} & 0.120 & 3.69 & -3 & 0.117$\pm$0.019 & -3.40$\pm$0.42 & 5.31$\pm$0.99 & 2.47e-05 & 4.06\\
\ion{Na}{I} & 0.126 & 3.40 & -2 & 0.139$\pm$0.015 & -4.38$\pm$0.54 & 10.07$\pm$1.26 & 3.96e-06 & 4.47\\
\ion{Mg}{I} & 0.084 & 3.33 & -3 & 0.062$\pm$0.005 & -8.40$\pm$1.40 & 33.45$\pm$3.30 & 3.51e-05 & 3.98\\\hline
\\
  \end{tabular}
  \caption{Summary of detected atomic species. Columns 1-4: Species name, peak relative absorption, signal-to-noise and location of the peak in the one-dimensional CCF. Columns 5-7: amplitude, centroid and \text{FWHM} of a Gaussian profile fit to the CCF feature, along with corresponding uncertainties. Columns 8 and 9: false-alarm-probabilities, and confidence level assuming Gaussianity.}
\label{tab:amptab}
\end{table*}

Our best-fit relative depth of \ion{Fe}{II} ($0.013\pm0.002\%$) is consistent with the $0.08\pm0.04\%$ depth of individual lines reported in combined HARPS-N data \citep{Casasayas2019} within $\lesssim2\sigma$. Likewise, our \ion{Na}{I} depth ($0.139\pm0.015\%$) is comparable to the previously reported $0.09\pm0.05\%$ combined strength of the Na D1 and D2 lines. In addition, all detected species have depths comparable to those found in the atmosphere of the ultra-hot Jupiter KELT-9 b\citep{Hoeijmakers2019}, but in contrast with KELT-9 b, this transit observation provides no evidence of absorption of \ion{Ti}{II}, \ion{Sc}{II} or \ion{Y}{II}. We have fixed $K_p$ at 175 \kms, consistent with the various optimal values of $K_p$ found by \citet{Casasayas2019,Nugroho2020,Stangret2020}. Variations of several \kms in $K_p$ do not change the recovered signal amplitudes or confidence levels appreciably.

To explain this discrepancy, we note that there is a difference in equilibrium tempeterature of almost 2,000 K between M2 ($T_{\textrm{eq}} \sim 2260$ K) and the much hotter KELT-9 b  ($T_{\textrm{eq}} \sim 4050$ K \citep{Gaudi2017}), so it is plausible that significant differences exist between the atmosphere chemistry and thermal structure of the atmospheres of these two planets. In addition, the present observations were obtained over the course of a single transit, whereas two transits of KELT-9 b were used by \citet{Hoeijmakers2019}. As a result, the combined cross-correlation functions have lower signal-to-noise than what was achieved in the analysis of KELT-9 b, resulting in lower sensitivities to trace species like \ion{Sc}{II} or \ion{Y}{II}, which were also least strongly detected in the sample of \citet{Hoeijmakers2019}.

With regards to the non-detection of \ion{Ti}{II} that was strongly detected in KELT-9 b, it is worth noting that the equilibrium temperature of M2 is slightly below that of WASP-121 b ($T_{\textrm{eq}} \sim 2358$ K) for which TiO condensation has been observed to be important \citep{Delrez2016,Evans2018}.

The peak locations of the CCF features generally suggest a blueshift of $\sim$ -1 to -9 \kms, which may be indicative of a day-to-night side wind or systematic errors in the determination of the systemic velocity, which can be difficult to constrain for fast-rotators \citep{Hoeijmakers2019}. However, recent works by \citet{Nugroho2020} and \citet{Stangret2020} also report blueshifts of the \ion{Fe}{I} and \ion{Fe}{II} signals of a few \kms, with \ion{Fe}{II} generally showing a smaller blueshift than \ion{Fe}{II}. Therefore, the interpretation that atmospheric absorption lines are significantly blue-shifted appears to be robust across multiple independent studies, indicating the presence of a day-to-night side wind. This wind appears to affect \ion{Fe}{I} and \ion{Fe}{II} differently, suggesting a stratification of the atmosphere, as alson hypothesised by \citet{Nugroho2020}.

The CCF for \ion{Ti}{II} exhibits a slight enhancement at 0.0 \kms\ with a S/N $\sim 2.4$. While this peak is the highest in the one-dimensional CCF, we refrain from claiming a detection. All detections listed in Table \ref{tab:amptab} have S/N of at least 3.0, and correspond to the maximum in the one-dimensional CCF. These two criteria were not met by the CCF of any other species.

The absorption signature of \ion{Mg}{I} appears significantly broader compared to the other species, with a measured FWHM of $33.45\pm3.30$ \kms versus $\sim 10$ \kms for the other species. On its own, the current data does not provide evidence to identify the cause of broadening processes that would act on \ion{Mg}{I} specifically. However, we note that diverse broadening has also been observed in the transmission spectrum of KELT-9 b, with \ion{Na}{I} and \ion{Mg}{I} showing FHWMs of $27.8 \pm 3.7$  and $27.5 \pm 4.3$ respectively \citep{Hoeijmakers2019} as opposed to values between 10 and 20 \kms for other species. Assuming that line broadening is mainly due to atmospheric dynamics, differences in the line-widths of different species would suggest that certain species may exist in distinct dynamical regimes in the upper atmospheres of ultra-hot Jupiters. This could include day-to-night side flows versus super-rotational jets \citep{showman2013,Louden2015,Brogi2016} or radial outflows \citep{Seidel2019}.  However, further analysis and observations will be needed to clarify the mechanism by which the absorption lines of certain species may be differentially broadened by these effects, and to what extent such processes are common among ultra-hot Jupiters.

\section{Discussion \& Conclusion}
\label{sec:con}

\subsection{Bootstrap and Temporal Variation}

The statistical treatment of the signatures as quoted above assumes that the noise in the cross-correlation function is normally distributed and uncorrelated. To determine the robustness of the detected signals, we determine false-alarm probabilities (FAP) using a bootstrap approach as follows. For a given species, we start with the two-dimensional, CCF time-series after removing the Doppler shadow. We randomly shuffle each row of the CCF, co-add in the rest-frame of the planet, fit a Gaussian profile at 0.0 \kms (i.e. the center of the randomly shuffled rows), and record the fitted amplitude. This process is repeated 50,000 times to populate a random distribution of amplitudes. When fitting the Gaussian, we enforce a $\text{FWHM} \geq 5$ \kms and a centroid $|\mu| \leq 20$ \kms. By setting a minimum \text{FWHM}, we require that a spurious CCF enhancement must be sufficiently wide to qualify as a false-alarm. Narrower absorption line profiles would be inconsistent with the minimum width set by the rotation of the planet, assuming tidal locking. Indeed, all of the reported detections in Table \ref{tab:amptab} have $\text{FWHM} > 5$ \kms. Finally, we fit the tail of the amplitude distribution with a power law model, and extrapolate it to the detected signal's strength. If we assume the distribution is Gaussian, the FAP can be converted to a $\sigma$-confidence level. The distribution of random amplitudes is shown in Fig. \ref{fig:fapfeii} for the case of \ion{Fe}{II}. Distributions for other species are in the Appendix.

The FAP does not account for autocorrelation between absorption lines from a given species, or spurious correlations between lines from different species. Since several CCFs show strong spurious signals (offset from the 0.0 \kms\ mark), we manually check the autocorrelation function of each model template. Additionally, we check the  cross-correlation function between each template and Fe I and Fe II, since these species contribute a multitude of strong absorption lines. The peak in the Mg I CCF at $+80$ \kms\ is due to spurious correlation with a strong nearby \ion{Fe}{II} line, also observed by \citet{Hoeijmakers2019}. \ion{Fe}{II} also produces a signal with \ion{Cr}{II} near $+50$ \kms. Autocorrelations for the detected species do not produce significant spurious signals, and spurious correlations with \ion{Fe}{I} are at much lower amplitude than the detections. Spurious features in CCFs of \ion{Na}{I} near $+90$ \kms, \ion{Cr}{II} near $+20$ \kms, and \ion{Ti}{II} near $+90$ \kms\ and $+170$ \kms cannot be easily explained, and may result from noise in the two-dimensional CCF.

\begin{figure}
  \centering
  \includegraphics[width=\linewidth,angle=0]{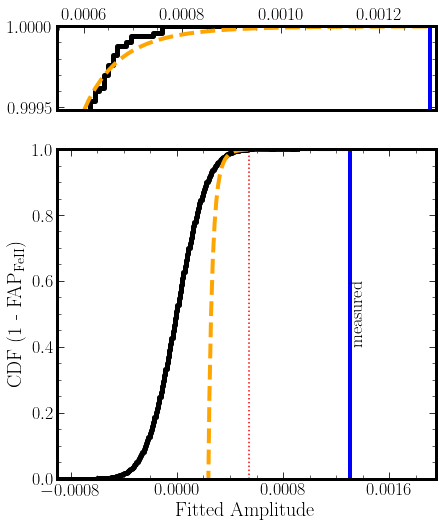}
      \caption{Cumulative Distribution Function (CDF) of amplitudes of Gaussian fits to the randomly shuffled and stacked \ion{Fe}{II} CCF. The black curve depicts the CDF. The blue solid line marks the strength of the \ion{Fe}{II} detection. The dotted red-line marks three standard deviations from the distribution mean, after which the distribution is fitted with a powerlaw, shown as a the orange dashed line. The upper panel zooms in on the tail of the CDF for clarity.}
\label{fig:fapfeii}
\end{figure}

\begin{figure}
   \centering
   \includegraphics[width=\hsize,angle=0]{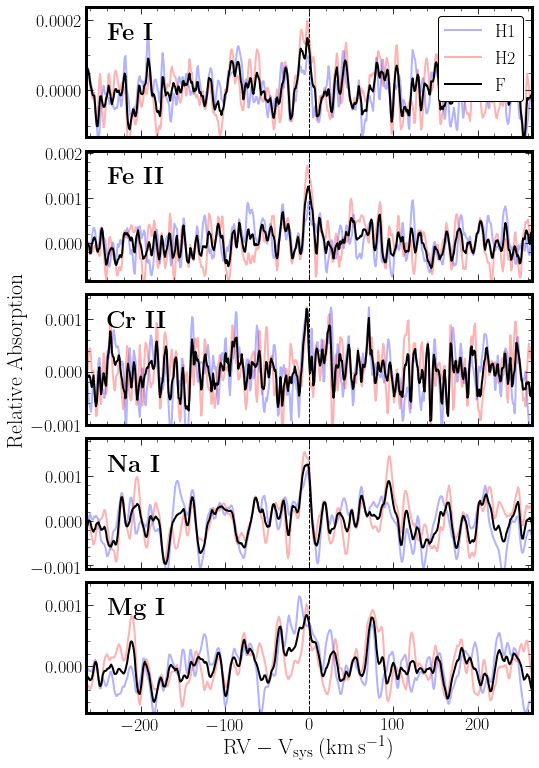}
      \caption{Comparison of co-added CCF features using data from the first half of transit (blue curve), second half (red curve) and full transit (black curve), for each of the strongest species detections (\ion{Fe}{I}, \ion{Fe}{II} and \ion{Cr}{II}). The legend denotes which in-transit exposures were used in the calculation (H1, H2, F denote first half of transit, second half of transit and full transit respectively). The dashed line indicates the expected location of the signal, at 0.0 \kms velocity offset.}
\label{fig:diff}
\end{figure}

\begin{figure}
  \centering
  \includegraphics[width=\linewidth,angle=0]{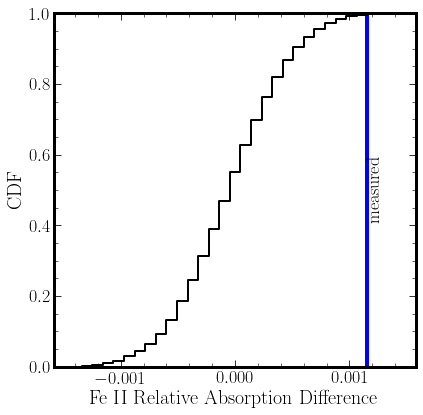}
      \caption{Cumulative Distribution Function of the difference in amplitudes recovered from injecting an artificial \ion{Fe}{II} signal, and recovering it in the first and second halves of transit separately. The vertical blue line marks the measured difference in amplitudes.}
\label{fig:bootstrap}
\end{figure}

The measured planetary absorption signal appears to be stronger in the second half of the transit than it does in the first half. This is most evident in the upper-right panel of Fig. \ref{fig:result}, where the signature of the \ion{Fe}{II} line appears visible by eye during the later exposures of the observing sequence. While this could be stochastic,  particularly because this analysis is based on a single transit, it may instead be a physical effect related to the distribution of \ion{Fe}{II} in the atmosphere of the planet, where it is present on the day-side near the evening (trailing) terminator that rotates into view towards the end of the transit, an effect that has recently been used to explain time-dependencies in the absorption spectrum of \ion{Fe}{I} in the atmospheres of WASP-121 b \citep{Bourrier2019} and WASP-76 b \citep{Ehrenreich2020}. We quantify this trend by fitting a Gaussian profile to the one-dimensional co-added CCF feature, using only the first and second halves of the in-transit exposures separately (Fig. \ref{fig:diff}). We take the difference between the two Gaussians' amplitudes of \ion{Fe}{II} (which shows the largest relative discrepancy between transit halves) and compare this to differences that might arise by chance, given a signal that is uniform throughout the transit.

First, we add a Gaussian absorption line of identical amplitude and \text{FWHM} as the detected absorption line to each row of the two-dimensional CCF, thereby injecting an artificial signal of identical strength. The centroid is set by the corresponding phase, and a randomly selected $0 < K_p < 300$ \kms, and $-300 < V_{\rm sys} < 300$ \kms\ (excluding $-20 < V_{\rm sys} < 20$ \kms to avoid overlap with the actual planet signal). We subsequently stack the two-dimensional CCF along the injected signal's velocities, treating the first and second halves separately. We fit new Gaussians at the injected signal's $V_{\rm sys}$ for each half, and record the difference in their amplitudes. We repeat this procedure 10,000 times, injecting and recovering a signal along random paths in the two-dimensional CCF. Results are shown in Figure \ref{fig:bootstrap}, where the observed difference or greater occurs in $1.2\times10^{-3}$ of all cases. Assuming the distribution is Gaussian, the observed \ion{Fe}{II} amplitude difference is at a 2.7$\sigma$ confidence level.

Variability of the depth of absorption features has been observed in the transmission spectra of KELT-9 b \citep{Cauley2019} and currently most clearly in WASP-76 b \citep{Ehrenreich2020}. In the case of WASP-76 b, this is attributed to differences in chemical composition between the morning (leading) and evening (trailing) terminators, due to the offset of the substellar hot-spot towards the evening terminator. These observations show the appearance of a significant blue-shift as the evening twilight region rotates into view, which is explained by the rotation of the planet in combination with a day-to-night-side wind: The transmission spectrum of the leading terminator probes cool gas that streams from the night-side towards the hot day-side. Conversely, the trailing limb contains hot gas that was strongly irradiated on the day-side and is approaching the night-side where it cools down. However, in the case of M2 b, the absorption signal appears to be symmetric around 0 \kms, which may indicate that the chemistry at both terminators is more similar than for WASP-76 b. An increase in the absorption line (which in this data may amount to a factor of 2, see Fig. 7), may however be indicative of a scale-height difference on both terminators, which could be caused by the temperature on the evening terminator being higher than on the morning terminator.

\subsection{Summary}
In this paper we present detections of atomic metal absorption lines in the transmission spectrum of the ultra-hot Jupiter MASCARA-2 b \citep{Lund2017,Talens2018}. To this end, one transit of MASCARA-2 b was observed with EXPRES, the high-resolution optical spectrograph newly commissioned at the Lowell Discovery Telescope. We confirm previous detections of \ion{Fe}{I}, \ion{Fe}{II} and \ion{Na}{I} \citep{Casasayas2018,Casasayas2019,Nugroho2020,Stangret2020}, and additionally find strong evidence for line absorption by atomic \ion{Mg}{I}, and \ion{Cr}{II}. All detected species appear to be blue-shifted, indicating the presence of a day-to-night side wind, also observed in previous studies \citep[e.g.][]{Nugroho2020,Stangret2020}. Using the shape variation of the stellar absorption lines induced by the transiting planet (i.e. the Doppler shadow), we constrain the projected spin-orbit misalignment to $1.6\pm3.1$ degrees, consistent with an aligned orbit. The cross-correlation functions indicate hints of time-variability in the absorption strength of these species, albeit at a level of $\lesssim 3 \sigma$. With a single transit, we cannot rule out that this variability is spurious. However, if it is astrophysical in origin, it potentially traces differential atmospheric structure between morning and evening terminators, reminiscent of what has recently been observed in the transmission spectrum of WASP-76 b \citep{Ehrenreich2020}. These results demonstrate the first spectroscopic observation of an exoplanet atmosphere with the EXPRES instrument, demonstrating its future potential for atmospheric characterisation.

\begin{acknowledgements}
This work was supported by the PlanetS National Centre of Competence in Research (NCCR) supported by the Swiss National Science Foundation (SNSF), the NSF under grants NSF MRI-1429365 and ATI-1509436 and by the European Re- search Council (ERC) under the European Union’s Horizon 2020 research and innovation programme (projects Four Aces and EXOKLEIN with grant agreement numbers 724427 and 771620, respectively). We acknowledge generous support for telescope time provided by the Heising-Simons Foundation and the Yale Astronomy Department. DAF and JMB wish to acknowledge support from an anonymous donation, which has also been used for telescope time. LLZ gratefully acknowledges support from the NSF GRFP. These results made use of the Lowell Discovery Telescope at Lowell Observatory. Lowell is a private, non-profit institution dedicated to astrophysical research and public appreciation of astronomy and operates the LDT in partnership with Boston University, the University of Maryland, the University of Toledo, Northern Arizona University and Yale University. We thank the Lowell Observatory astronomers and staff for their extraordinary support.
\end{acknowledgements}

\bibliographystyle{aa}
\bibliography{main}

\begin{appendix}
\section{MCMC Results, Additional Figures}
\pagebreak

\begin{table*}
  \centering
  \begin{tabular}{l | c c r r r}
\hline
\hline
Parameter & Symbol & Unit & MCMC Median & T18 (Prior) & L17 \\
\hline
Scaled Semi-major Axis           & $a/R_*$	      & -        & $7.50 \pm 0.04$              & $7.50 \pm 0.04$      & $7.44^{+0.14}_{-0.13}$ \\
Projected Obliquity              & $\lambda$      & $^\circ$ & $1.58^{+3.14}_{-3.12}$       & $0.6 \pm 4$          & $3.4 \pm 2.1$ \\
Orbit Inclination                & $i_p$          & $^\circ$ & $86.16 \pm 0.5$              & $86.4^{+0.5}_{-0.4}$ & $86.15^{+0.28}_{-0.27}$ \\
Projected Stellar Rotation Speed & $v\sin i_*$    & \kms     & $113.50^{+0.77}_{-0.74}$     & $114 \pm 3$          & $115.9 \pm 3.4$ \\
RV Offset                        & -              & \kms     & $0.7 \pm 3.1$                & $0 \pm 5^*$ & - \\
\hline
  \end{tabular}
  \caption{MCMC results for modelling of the Rossiter-McLaughlin effect. The model consists of the four parameters listed in column one, plus a global RV offset. The MCMC median values are listed in column four. Uncertainties correspond to the 16$^{\rm th}$ and 84$^{\rm th}$ percentiles. For comparison, we list literature values in column five, and their references in column 6. T18 denotes \citet{Talens2018}, and L17 denotes \citet{Lund2017}. Asterisk ($^*$) indicates RV Offset prior not from literature.}
\label{tab:rmtab}
\end{table*}

\begin{figure*}
    \centering
   \includegraphics[width=1.0\linewidth]{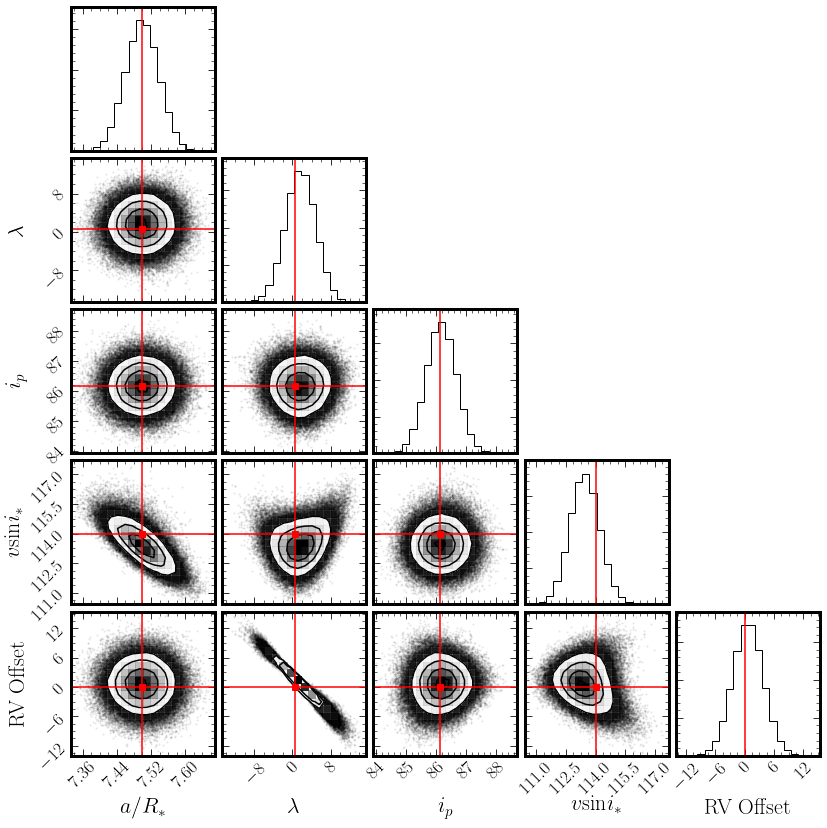}
      \caption{Correlation diagram from our MCMC fit to the observed Doppler shadow, as well as one-dimensional histograms. The model consists of five parameters:  scaled semi-major axis ($a/R_*$), projected obliquity ($\lambda$), orbital inclination ($i_p$), projected stellar rotation speed ($v\sin i_*$), and a global RV offset. Red lines denote literature values for the parameters \citep{Talens2018}.}
         \label{fig:rmmcmc}
   \end{figure*}

\begin{figure*}
    \centering
   \includegraphics[width=12cm,angle=0]{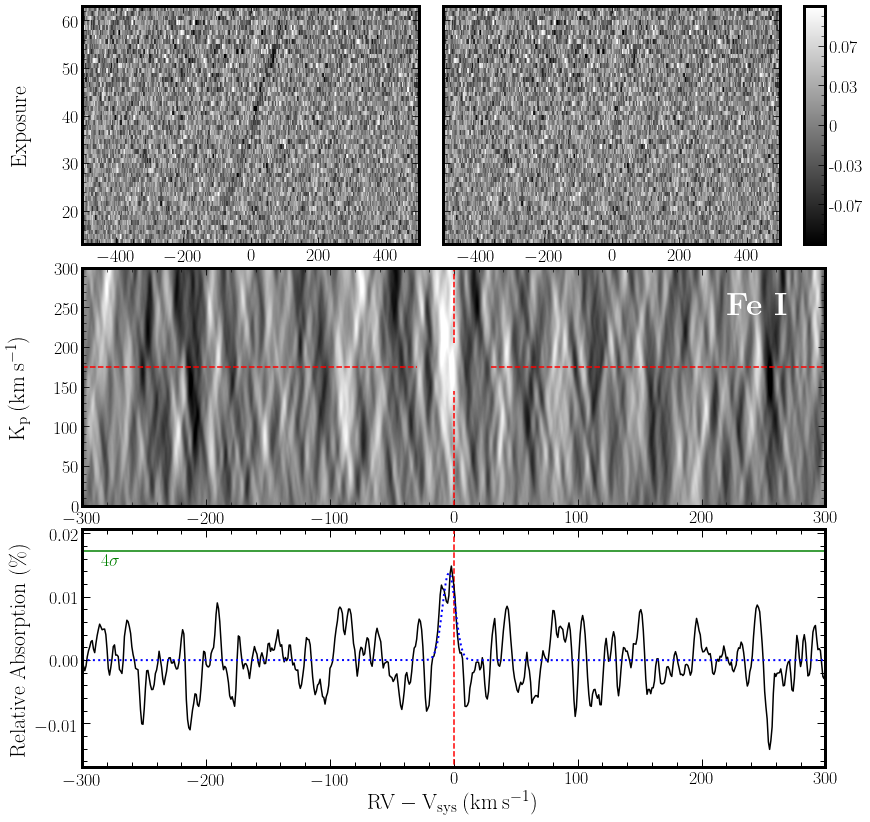}
      \caption{Same as Figure \ref{fig:result}, depicting the cross-correlation process for \ion{Fe}{I}. CCFs are in the rest-frame of the planet.}
\end{figure*}

\begin{figure*}
    \centering
   \includegraphics[width=12cm,angle=0]{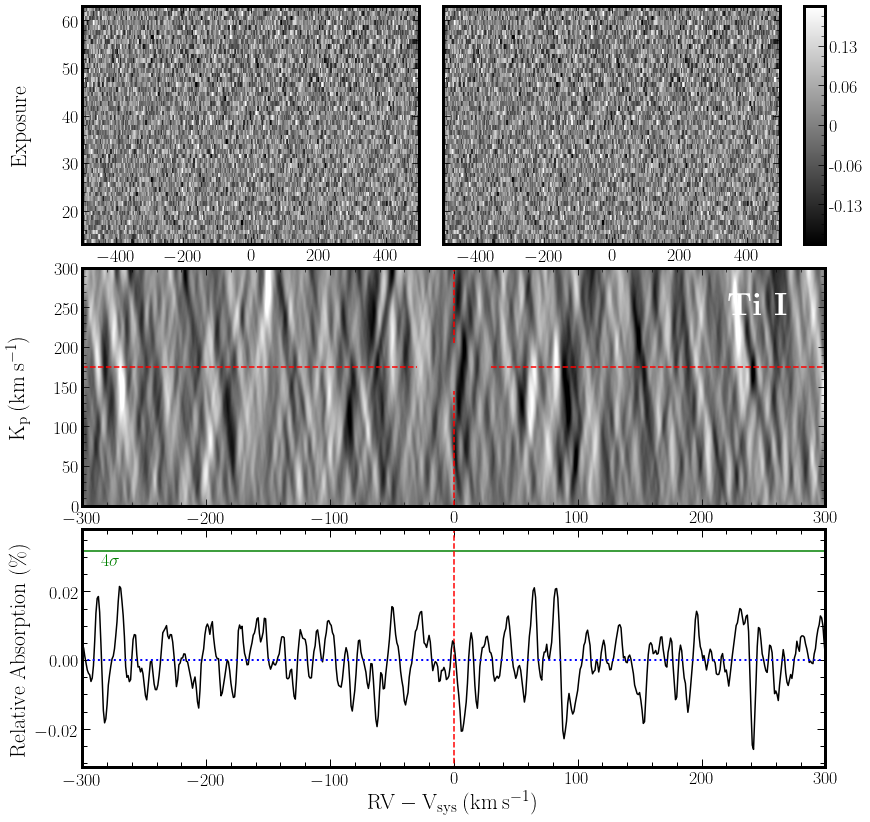}
      \caption{Same as Figure \ref{fig:result}, depicting the cross-correlation process for \ion{Ti}{I}. CCFs are in the rest-frame of the planet.}
\end{figure*}

\begin{figure*}
    \centering
   \includegraphics[width=12cm,angle=0]{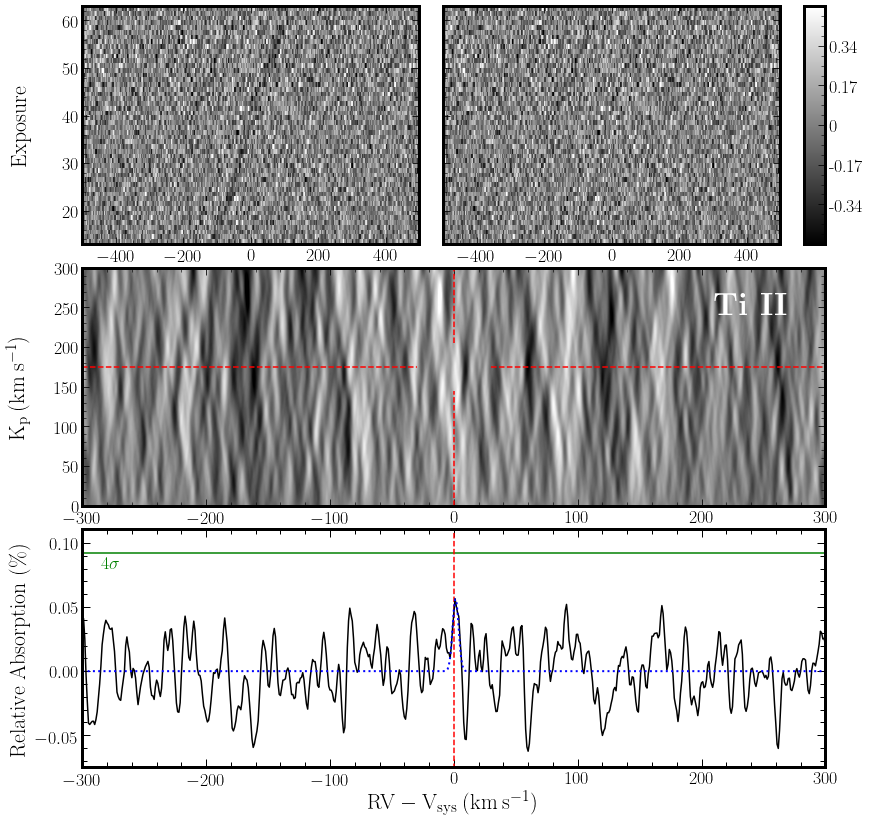}
      \caption{Same as Figure \ref{fig:result}, depicting the cross-correlation process for \ion{Ti}{II}. CCFs are in the rest-frame of the planet.}
\end{figure*}

\begin{figure*}
    \centering
     \includegraphics[width=12cm,angle=0]{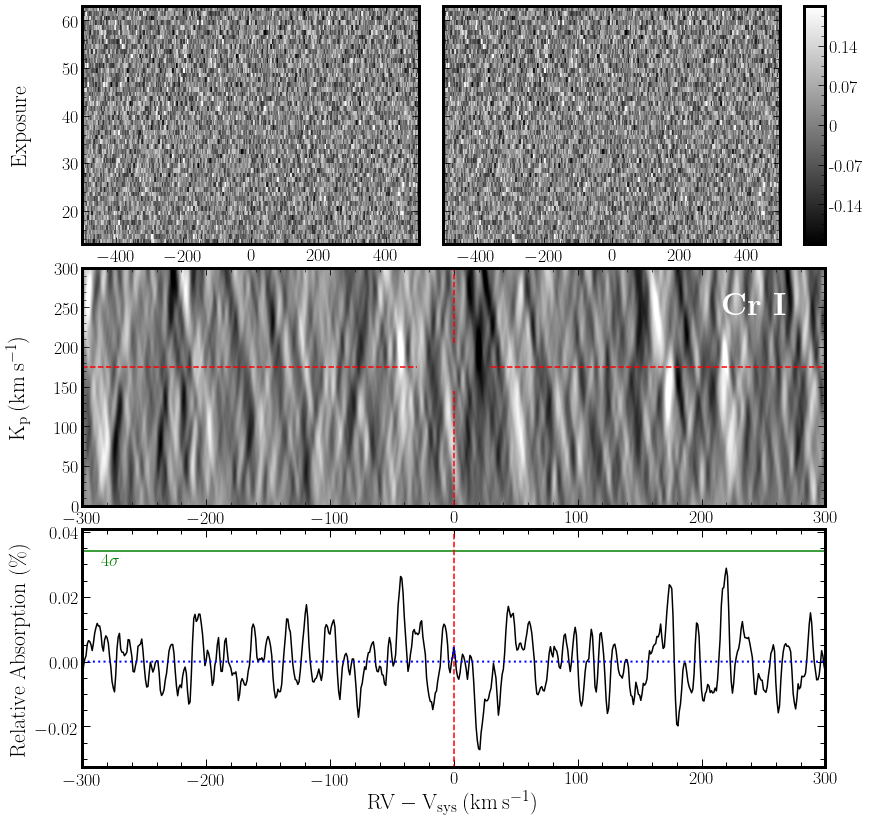}
      \caption{Same as Figure \ref{fig:result}, depicting the cross-correlation process for \ion{Cr}{I}. CCFs are in the rest-frame of the planet.}\end{figure*}

\begin{figure*}
    \centering
   \includegraphics[width=12cm,angle=0]{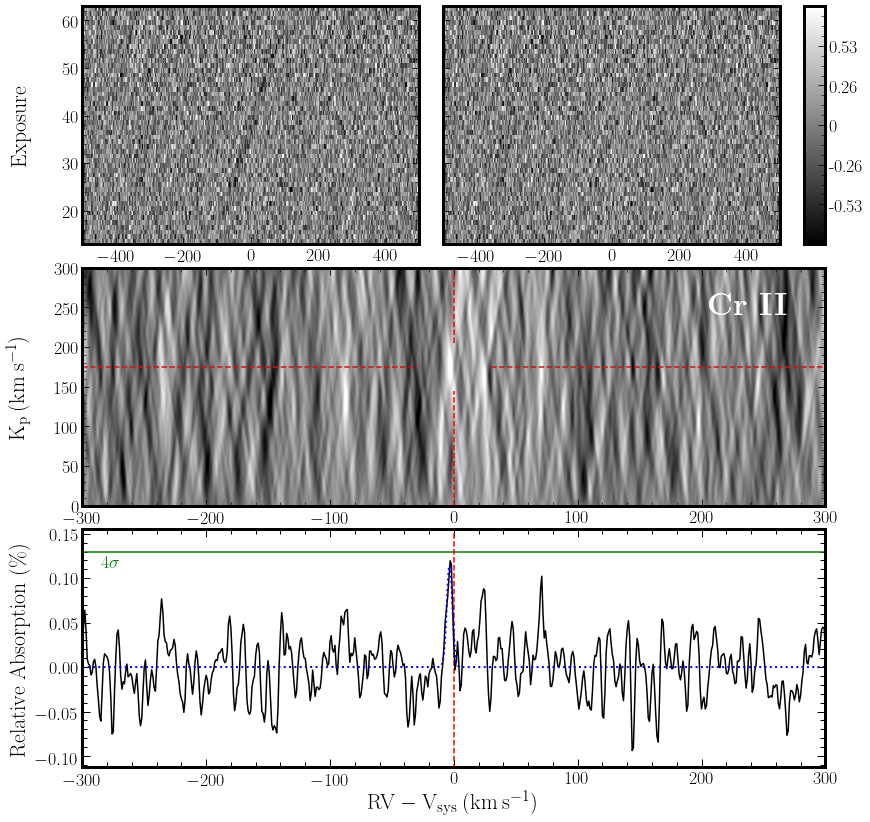}
      \caption{Same as Figure \ref{fig:result}, depicting the cross-correlation process for \ion{Cr}{II}. CCFs are in the rest-frame of the planet.}\end{figure*}

\begin{figure*}
    \centering
  \includegraphics[width=12cm,angle=0]{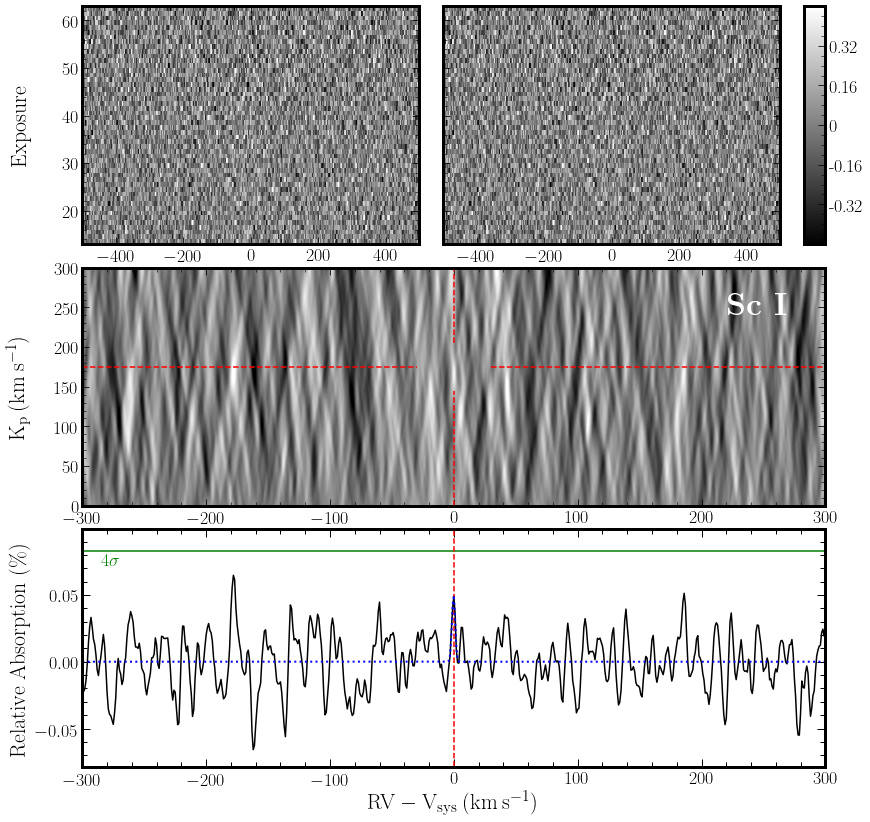}
      \caption{Same as Figure \ref{fig:result}, depicting the cross-correlation process for \ion{Sc}{I}. CCFs are in the rest-frame of the planet.}
\end{figure*}

\begin{figure*}
    \centering
   \includegraphics[width=12cm,angle=0]{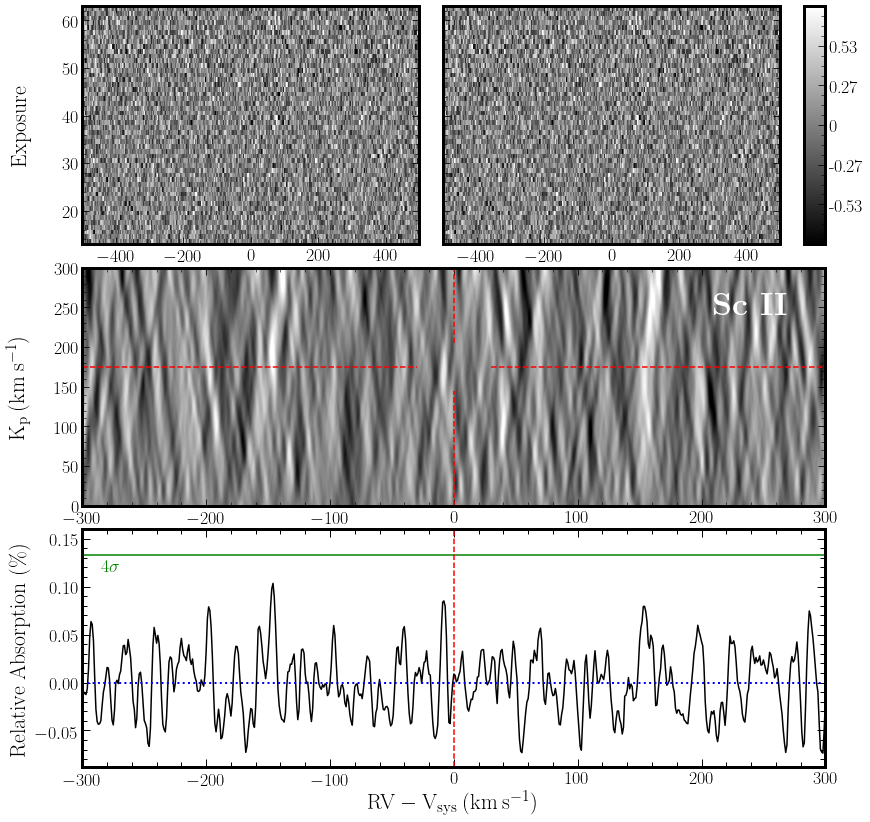}
      \caption{Same as Figure \ref{fig:result}, depicting the cross-correlation process for \ion{Sc}{II}. CCFs are in the rest-frame of the planet.}\end{figure*}

\begin{figure*}
    \centering
    \includegraphics[width=12cm,angle=0]{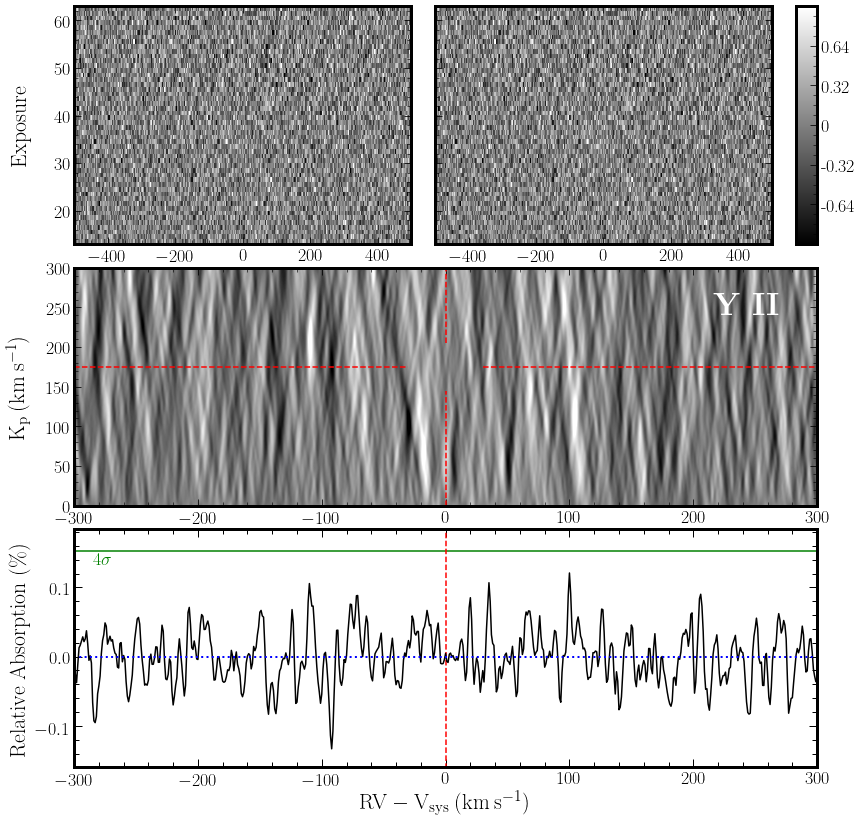}
      \caption{Same as Figure \ref{fig:result}, depicting the cross-correlation process for \ion{Y}{II}. CCFs are in the rest-frame of the planet.}\end{figure*}

\begin{figure*}
    \centering
   \includegraphics[width=12cm,angle=0]{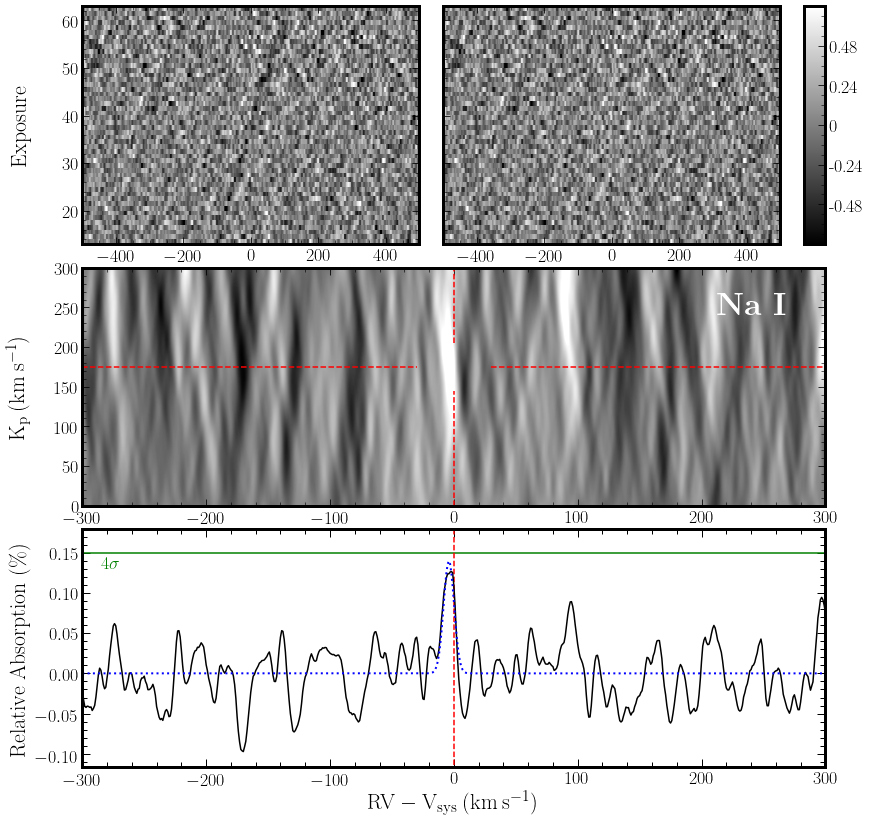}
      \caption{Same as Figure \ref{fig:result}, depicting the cross-correlation process for \ion{Na}{I}. CCFs are in the rest-frame of the planet.}\end{figure*}

\begin{figure*}
    \centering
   \includegraphics[width=12cm,angle=0]{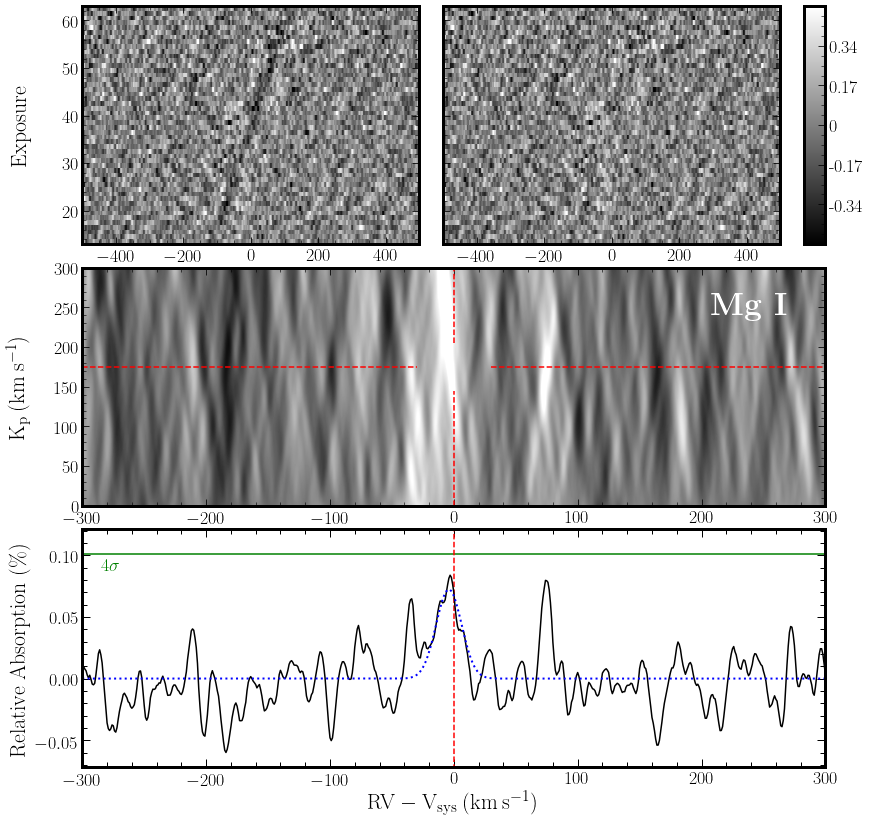}
      \caption{Same as Figure \ref{fig:result}, depicting the cross-correlation process for \ion{Mg}{I}. CCFs are in the rest-frame of the planet.}
\end{figure*}

\begin{figure*}
    \centering
   \includegraphics[width=9cm,angle=0]{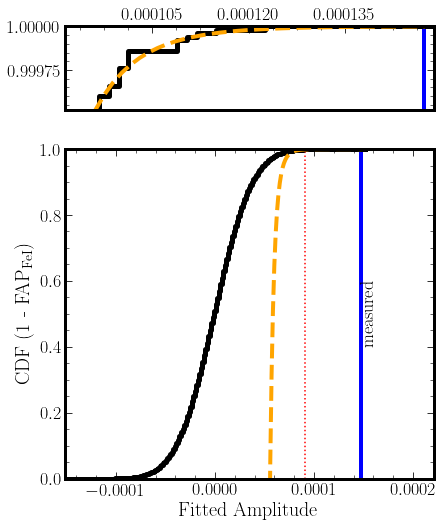}
      \caption{Same as Fig \ref{fig:fapfeii}, except based on cross-correlation with the \ion{Fe}{I} template. This plot shows the distribution of strengths of random signals generated by the CCFs, from which the FAP is derived.}
\end{figure*}

\begin{figure*}
    \centering
   \includegraphics[width=9cm,angle=0]{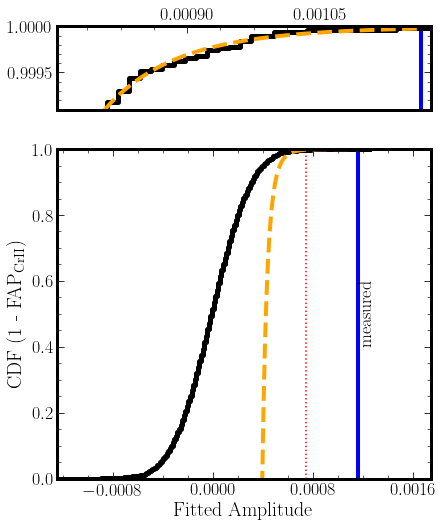}
      \caption{Same as Fig \ref{fig:fapfeii}, except based on cross-correlation with the \ion{Cr}{II} template. This plot shows the distribution of strengths of random signals generated by the CCFs, from which the FAP is derived.}
\end{figure*}

\begin{figure*}
    \centering
   \includegraphics[width=9cm,angle=0]{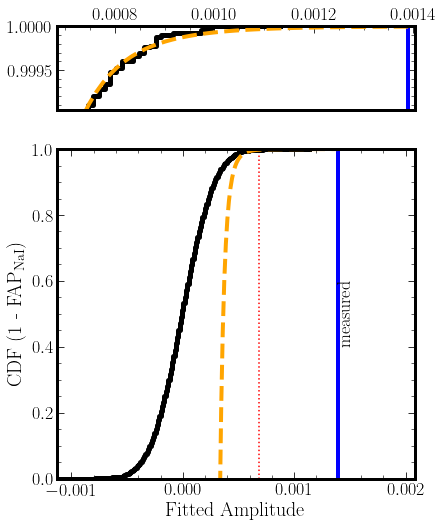}
      \caption{Same as Fig \ref{fig:fapfeii}, except based on cross-correlation with the \ion{Na}{I} template. This plot shows the distribution of strengths of random signals generated by the CCFs, from which the FAP is derived.}
\end{figure*}

\begin{figure*}
    \centering
   \includegraphics[width=9cm,angle=0]{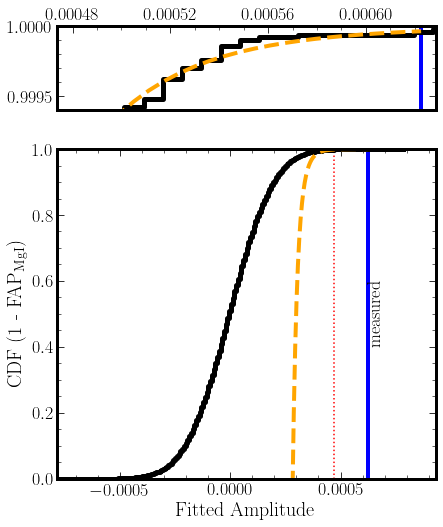}
      \caption{Same as Fig \ref{fig:fapfeii}, except based on cross-correlation with the \ion{Mg}{I} template. This plot shows the distribution of strengths of random signals generated by the CCFs, from which the FAP is derived.}
\end{figure*}


\end{appendix}

\end{document}